# “Delving into” the quantification of Ai-generated content on the internet (synthetic data)

Dirk HR Spennemann [1,2]*

[1] Gulbali Institute; Charles Sturt University; Albury NSW 2640, Australia
[2] School of Agricultural, Environmental and Veterinary Sciences, Charles Sturt University; Albury NSW 2640, Australia
* Correspondence: dspennemann@csu.edu.au

## Abstract

While it is increasingly evident that the internet is becoming saturated with content created by generated Ai large language models, accurately measuring the scale of this phenomenon has proven challenging. By analyzing the frequency of specific keywords commonly used by ChatGPT, this paper demonstrates that such linguistic markers can effectively be used to estimate the presence of generative AI content online. The findings suggest that at least 30% of text on active web pages originates from AI-generated sources, with the actual proportion likely approaching 40%. Given the implications of autophagous loops, this is a sobering realization.

## Introduction

OpenAi has released a number of Generative Pre-trained Transformer (GPT) models, with the early versions of GPT-1 (2018) [1], GPT-2 (November 5, 2019) and GPT-3 (May 2020) [2].ChatGPT3.5 was launched on 30 November 2022 and was the first model accessible by the wider public [3]. It has seen a range of significant improvements with a continued model roll out of GPT-4 (March 14, 2023)[4], GPT-4o (‘omni’) (May–August 2024)[5], GPT-4o mini (July 18, 2024) and the current version GPT4.5 (February 23, 2025) [6].

The training data used for the development of generative Ai large language models, such as ChatGPT and DeepSeek, are derived from fiction and non-fiction books, government documents, articles, and web pages to establish the parameters of language, while a considerable amount of factual knowledge has been taken from the Common Crawl set of 250 billion pages scraped off the internet [7,8], as well as from sites such as Wikipedia [9]. The training and red hatting phase ensured the semantic integrity of responses, the general accuracy, the emotional valence and effectiveness of filters to exclude harmful responses ®. Despite testing and subsequent moderation activities, biases derived from the training data remain ®[7,10].

The dramatic uptake of generative Ai by writers and contributors to traditional as well as social media, often with little editing or modification of the generated text, has resulted in a proliferation of webpages that contain or solely consist of Ai generated text [11]. Some generative Ai applications have the ability to search the internet and, purportedly, add that information to the data available for interpretation and analysis incorporating the information in their responses to human prompts ®. This has led to concerns that that web-enabled generative Ai will not only draw on human-generated base information (as per training data), but will also use ‘synthetic data’, i.e. information that was aggregated and generated by generative Ai itself [12]. Colloquially labelled ‘A’ cannibalism,’ there is the justifiable fear that a recursive (‘autophagous’) loop will develop where the human-generated original content accessible generative Ai becomes increasingly diluted by generative Ai-created, recombined text [13,14]. While synthetic data may be used in model training [15], autophagous loops lead to linguistic entrophy [16] and will be detrimental with regard to factual information, in particular where the content of web pages purveys extreme positions or where the web has been flooded with similar messaging by malevolent actors trying to bias public perception [17]. As early as May 2023 over fifty news websites could be identified as 'news copy' which was entirely Ai generated [18]. By February 2025 that had risen to at least 1,254 [11] churning out copy on a daily basis.

While the presence and potential implications of generative Ai created synthetic data in the interact has been widely discussed, there are only few data that show the possible extent of the phenomenon. Some data are based on algorithms that purport to be able to distinguish between text written by humans or generative Ai. A study by Copyleaks stated that 1.57% of a sample of 1 million pages were Ai generated [19].

This paper provides an attempt at quantifying the extent of synthetic data on the internet, but takes a different approach. It finds in roots in the observation that many of the conversations documented as part of datasets for various papers examining the abilities, biases and limitations of ChatGPT contained the phrase 'delve into' as part of the generative Ai response, (e.g. [20,21]).

## Methodology

Monthly frequencies of webpages that contain specific phrases were derived from a simple Google search for term ['delve into', | 'explore'] with the search period time-restricted to a monthly interval. This ensures webpages that were created or revised (with a new date) during that month. Monthly data were collected for the period January 2020 to March 2025. The annual frequency of scholarly citations using the phrase 'delve into' was derived from time-restricted searches in GoogleScholar. The frequency of world-wide Google searches was derived from weekly data provided by Google Trends.[22,23]

## Results and Discussion

The monthly frequencies of webpages containing the phrase 'delve into' remained stable from January 2020 to March 2023 (Figure 1), after which they began to rise steeply until January 2024. February 2024 represents a spike more than doubling the number of occurrences of the previous month. The figures drop back in the next month and further in April 2024, but remain on a linear trajectory from April 2023 to January 2025 when they begin to ease off. The frequency curve is punctuated by a series of troughs which are artefacts of Google indexing and do not represent genuine declines.

The time lag between the public release of ChatGPT on 30 November 2022 and dramatic increase in webpages containing the phrase 'delve into' from April 2023 onwards represents the experimentation phase of new technology leading to eventual adoption. Data derived from Google Trends show that web searches for ChatGPT rose steeply from November 2022 to February 2023, when it slowed (Figure 5). Data on the development of the user base of ChatGPT are patchy, but suggest a swift uptake, with the one million user mark being passed five days after the public launch [24]. By the end of January 2023 it had reached the 100 million user mark [25] rising to 180.5 million in March 2024 [24].

The predilection for ChatGPT to 'delve into' topics has decreased with the rollout of ChatGPT4o May–August 2024 and, more recently, with ChatGPT4.5 released in February 2025, which is borne out by the data (Figure 1). Anecdotally, the phraseology of 'delve into' seems to have been replaced by 'explore.' To assess this, the frequency of mentions of the word 'explore' in web pages shows a semi-regular pattern where some months are significantly lower than the adjacent months (Figure 2). This is an artefact derived from Google indexing of webpages as the same pattern can be observed when examining the frequency of mentions of the phrase 'delve into' (Figure 6). This then suggests that the observed troughs in the frequency distribution of the phrase 'delve into' (as observed for the months of June and October 2023 as well as February, May, September 2024 and January and February 2025) are also artefacts and that the genuine frequency for these months rests somewhere between the two adjacent values. Moreover, the spike in the frequency of the phrase 'delve into' in February 2024 corresponds with a trough in the frequency of the word 'explore' suggesting that the actual frequency of the phrase 'delve into' is much greater than reported.

At this point a word of caution needs to be inserted. It can be posited that the repeated exposure to the phrase 'delve into' will have led a number of people to incorporate that phrase into their vocabulary and that, therefore, at least some of the usage on webpages stems from this rather than copying of ChatGPT-generated text. This is, however, difficult to quantify. We can draw on academic publications as a proxy, under the assumption that, by and large, academics are unlikely to uncritically adopt full sections of generative Ai created text., especially since publishers moved quickly to delegitimise this (Committee on Publication Ethics, 2023; Formosa et al., 2024). The annual frequency of academic papers containing the phrase 'delve into' shows a linear increase between 2010 and 2022 with a high correlation coefficient (Pearson's $r=0.9961$, $df=11$, $p< 0.00001$) (Figure 4). Based on the regression formula for that distribution ($y = -2815508 + 1408 \cdot x$), we can calculate the 'natural' progressive increase for the years 2023 and 2024, demonstrating an increase of 27.3% for 2023 and 28.1% for 2024. As the 2025 data only encompass a single quarter, any projections for the year are too uncertain.

Even when assuming that the proportion of the general public incorporating the phrase 'delve into' their vocabulary is three times that of academics, the overall impact of that adjustment is only small. Cumulatively, for the period April 2023 to March 2025, a total of 176.5 million web pages were added that contain the phrase 'delve into.' The figure rises to 190.8 million if the troughs are smoothed out. When adjusted for the incorporation of the phrase into the people's general vocabulary, the figure is reduced to 167.9 (smoothed 186.1) million pages (to 151.1 [150.7] million if 3x academics were to be assumed).

Turning to the word 'explore' and adjusting the monthly figures of the smoothed data (troughs removed by averaging adjacent years), by applying the concept of the underlying baseline (see red trendline, Figure 7), then the 'net gain' for the period April 2023 to March 2025 equates to 3.396 billion pages. As 'explore' is a word in common parlance, there is no need to further adjust the figures—if anything, the usage of 'explore' has declined in academic publications in 2023 and 2024.

Extrapolating the proportion of webpages with Ai-generated content on the current internet is fraught with problems as accurate data on active webpages do not exist. Some approximation can be made, however. As of 20 March 2025, the internet comprised of 1.18 billion websites, 196.8 million of which were active [26]. The average number of indexed web pages is about 45 billion [27]. Applying the ratio of active vs legacy sites to web pages, then there are approximately 8.3 billion 'active' pages.

Thus, the 167.9 million pages (adjusted) that contain the phrase 'delve into' represent 2.2% of the total active web pages. The figure changes when we also consider 'explore'. The 3.396 billion pages created between April 2023 to March 2025 that contain the word 'explore' equate to 40.9% of all currently indexed (and active) web pages. As a caveat it should be noted that a fair number of web pages may well explore the functionality of generative Ai and use that term in their text, implying that the 40% figure is in all likelihood somewhat inflated. The extent of this is unclear. Even if we were to assume that the usage of 'explore' in a quarter of the post ChatGPT webpages is innocuous and unrelated (although that defies the long-term trend), the fact then remains that more than 30% of the current pages on the web are 'contaminated' by synthetic data.

## Conclusions

While there can be little doubt that the internet is increasingly 'polluted' by texts derived from generative Ai, the extent of this has been difficult to quantify. Using frequencies of selected keywords, this paper demonstrated that some words or phrases, favoured by ChatGPT in its responses, can be fruitfully deployed to provide an estimate of generative Ai text on the internet. The paper posits that at least 30% of the texts on active webpages is derived from generated Ai, with the figure more likely to be closer to 40%. In light of the implications discussed in the introduction, this is a sobering thought.

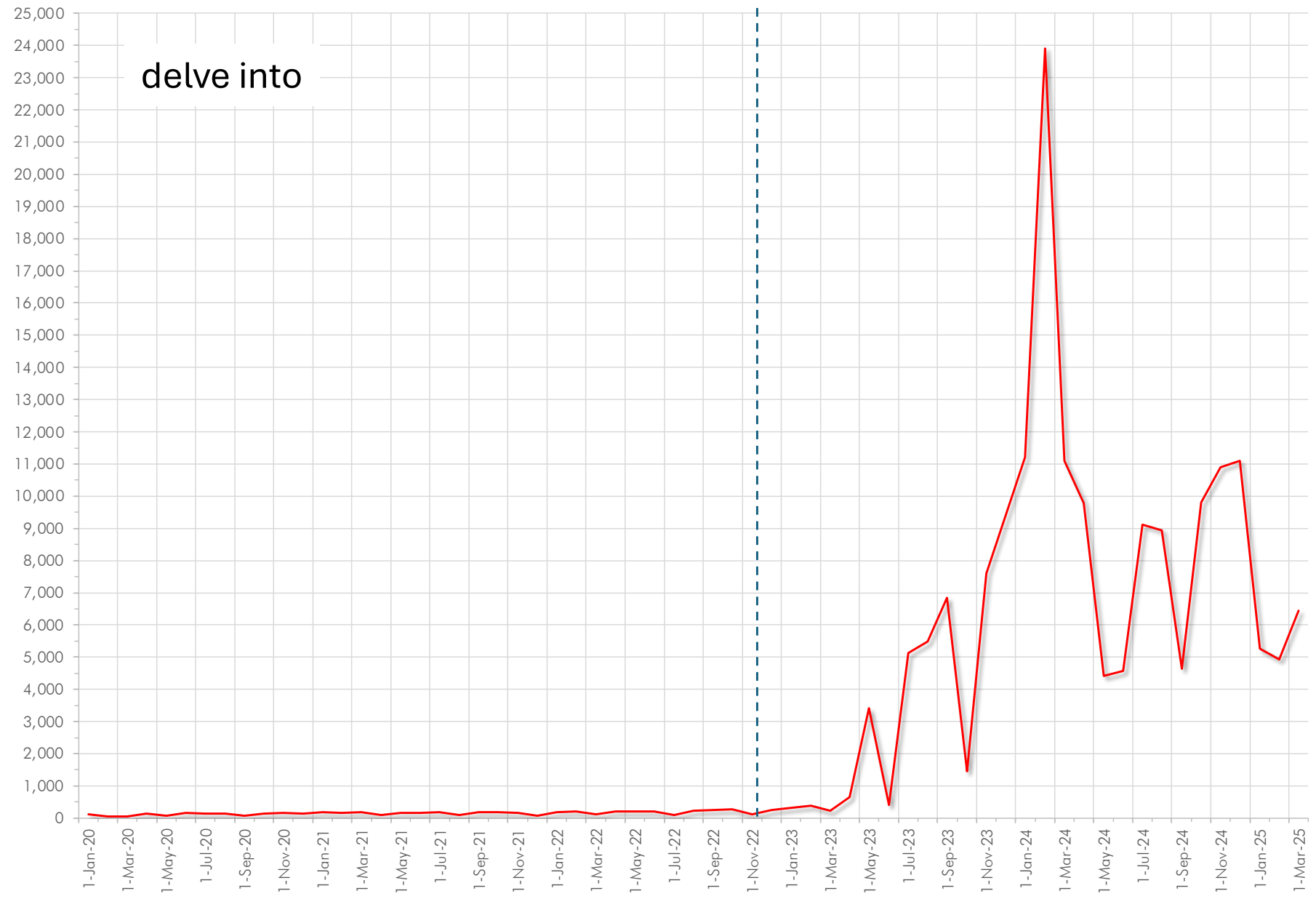

*Figure 1. Monthly frequency of webpages that contain the phrase 'delve into' (in 1,000) for the period January 2020 to March 2025 The dashed line indicated the public release of ChatGPT in November 2022.*

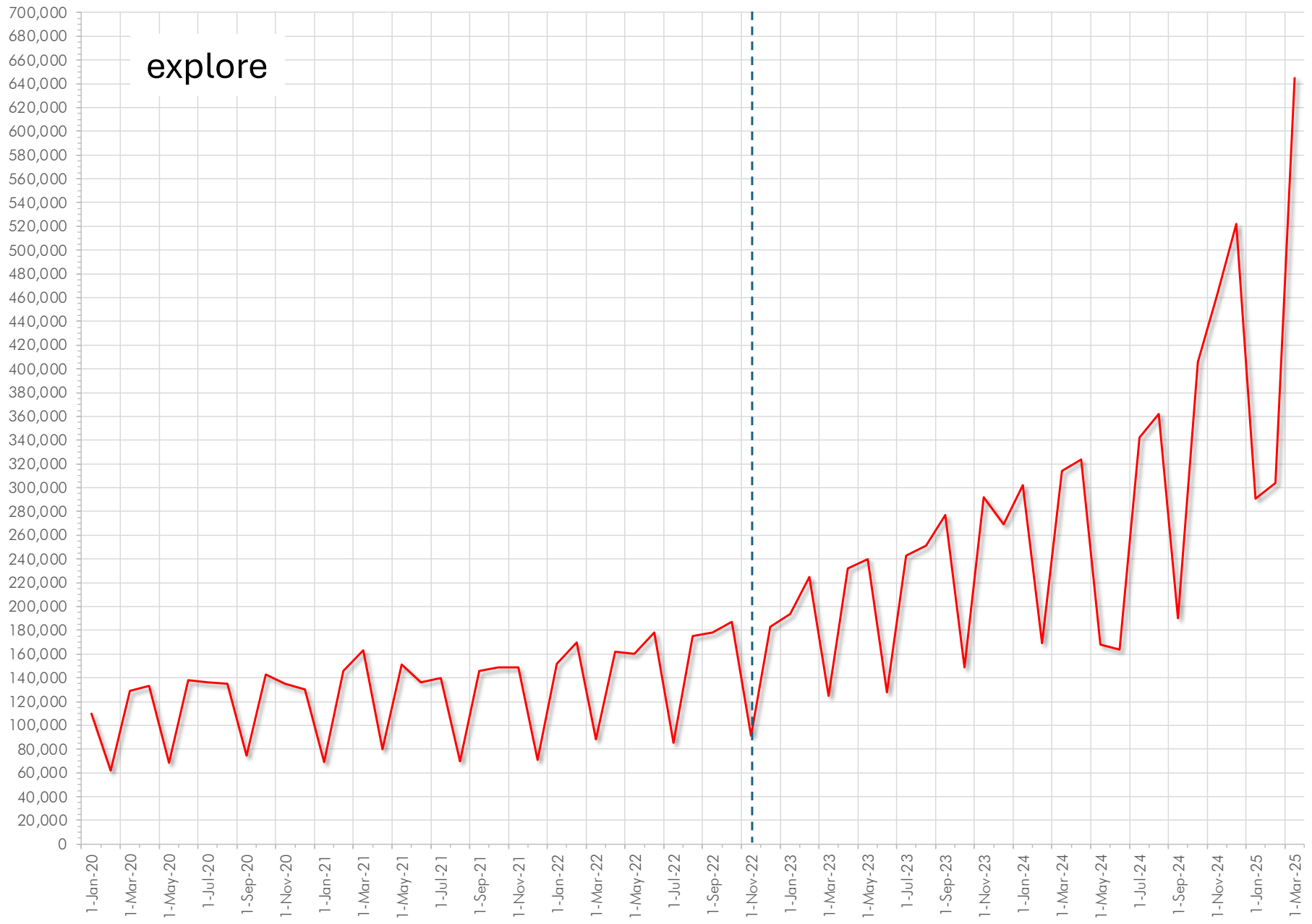

*Figure 2. Monthly frequency of webpages that contain the word 'explore' (in 1,000 for the period January 2020 to March 2025). The dashed line indicated the public release of ChatGPT in November 2022*

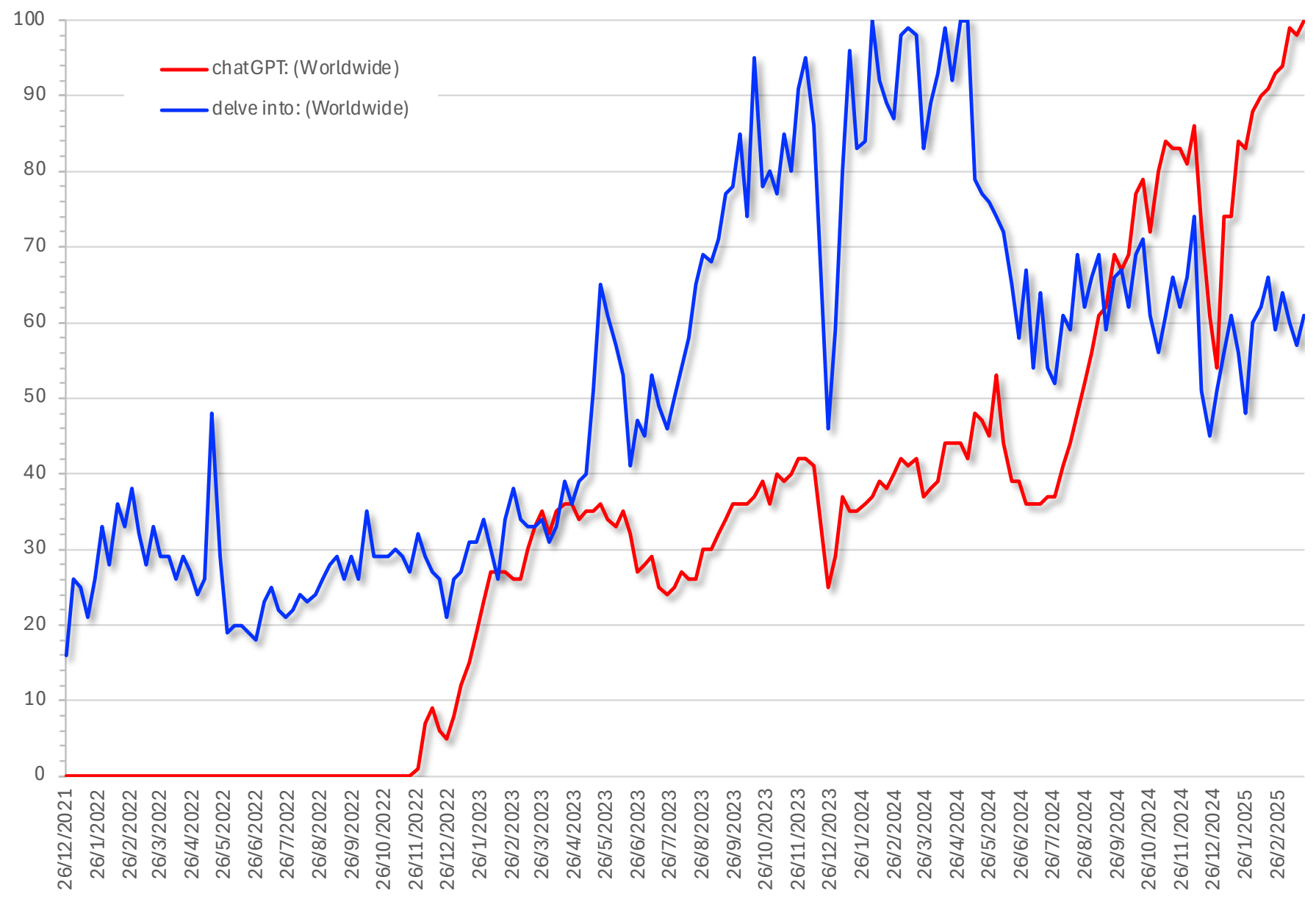

*Figure 3. Weekly frequency of world-wide Google searches (source Google Trends)*[22,23]

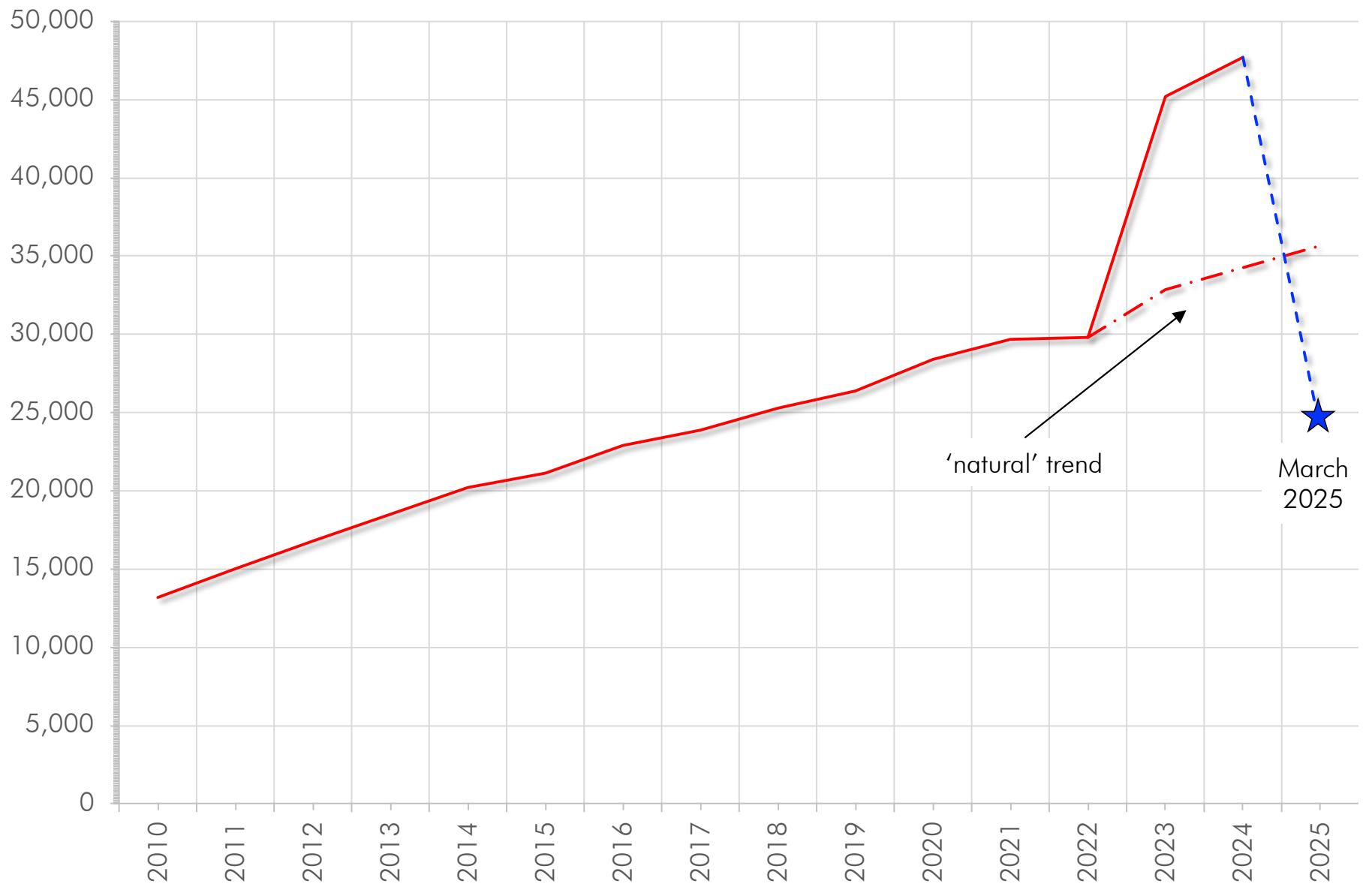

*Figure 4. Annual frequency of academic papers that contain the phrase 'delve into', 2010–March 2025.*

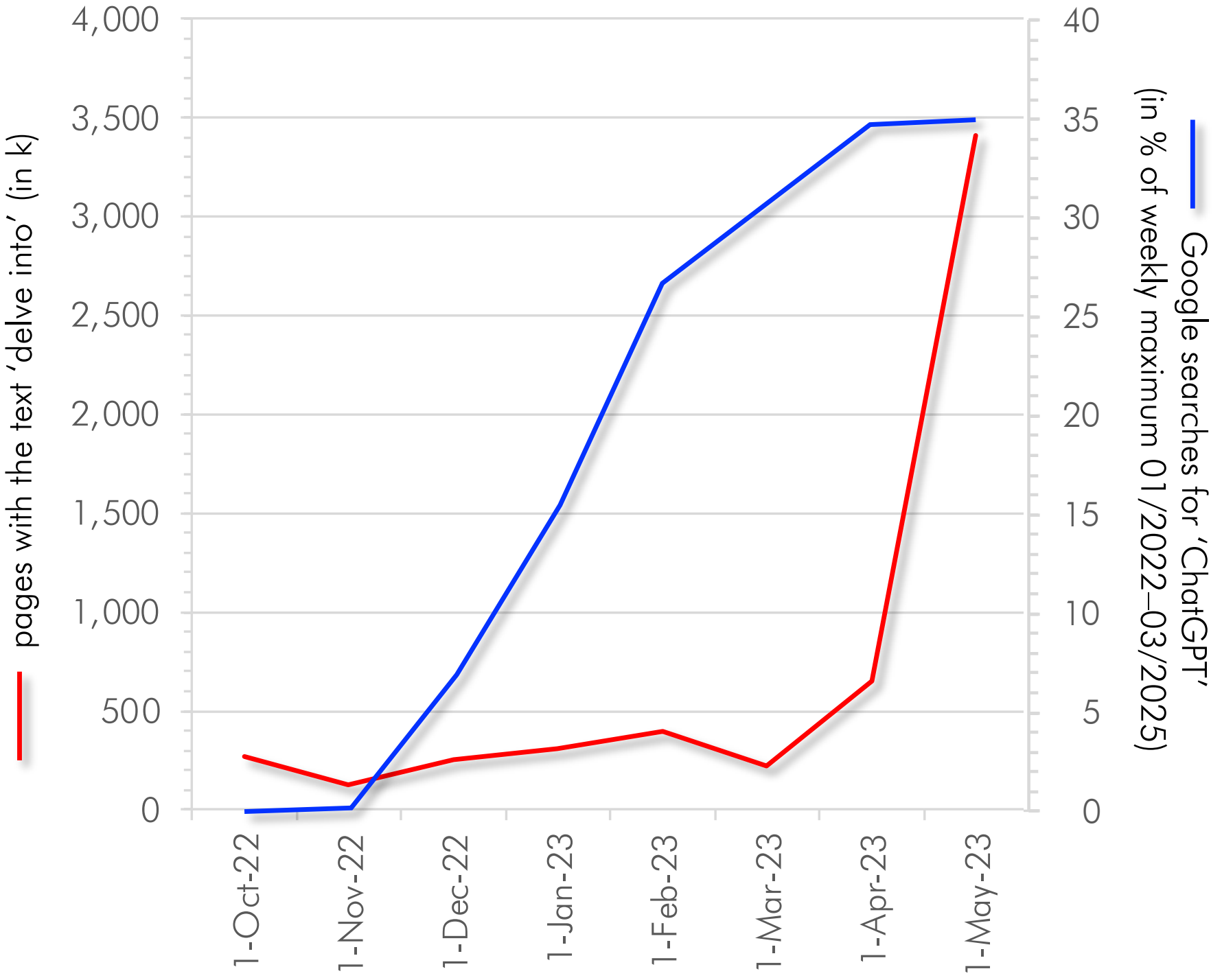

*Figure 5. Monthly frequencies of webpages that contain the phrase 'delve into' (in 1,000) and of world-wide Google searches for the term 'ChatGPT' for the period October 2022 to May 2023.*

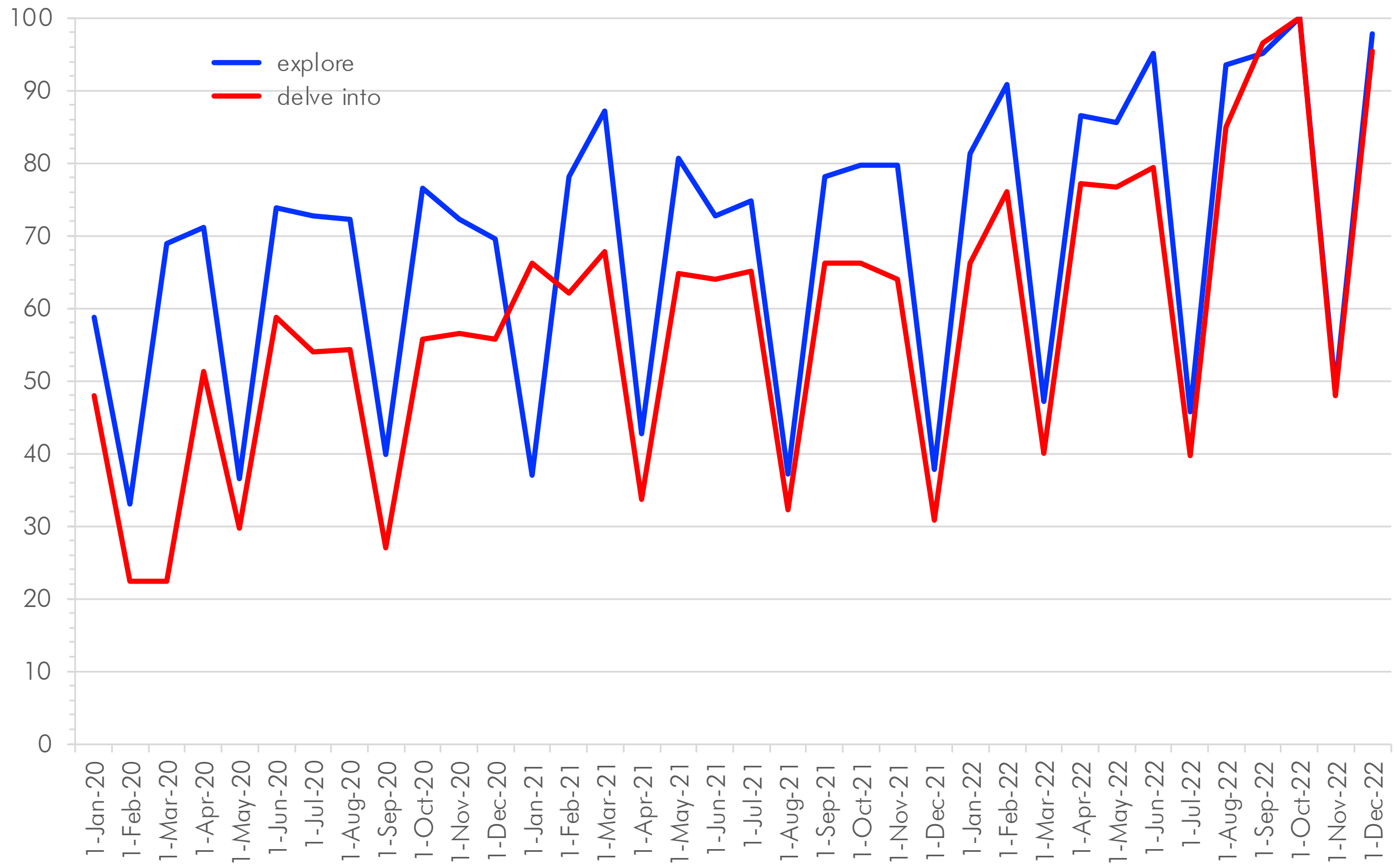

*Figure 6. Monthly frequencies of webpages that contain the phrases 'delve into' or 'explore' for the period January 2020 to December 2022 (expressed in % of the monthly maximum value for the period).*

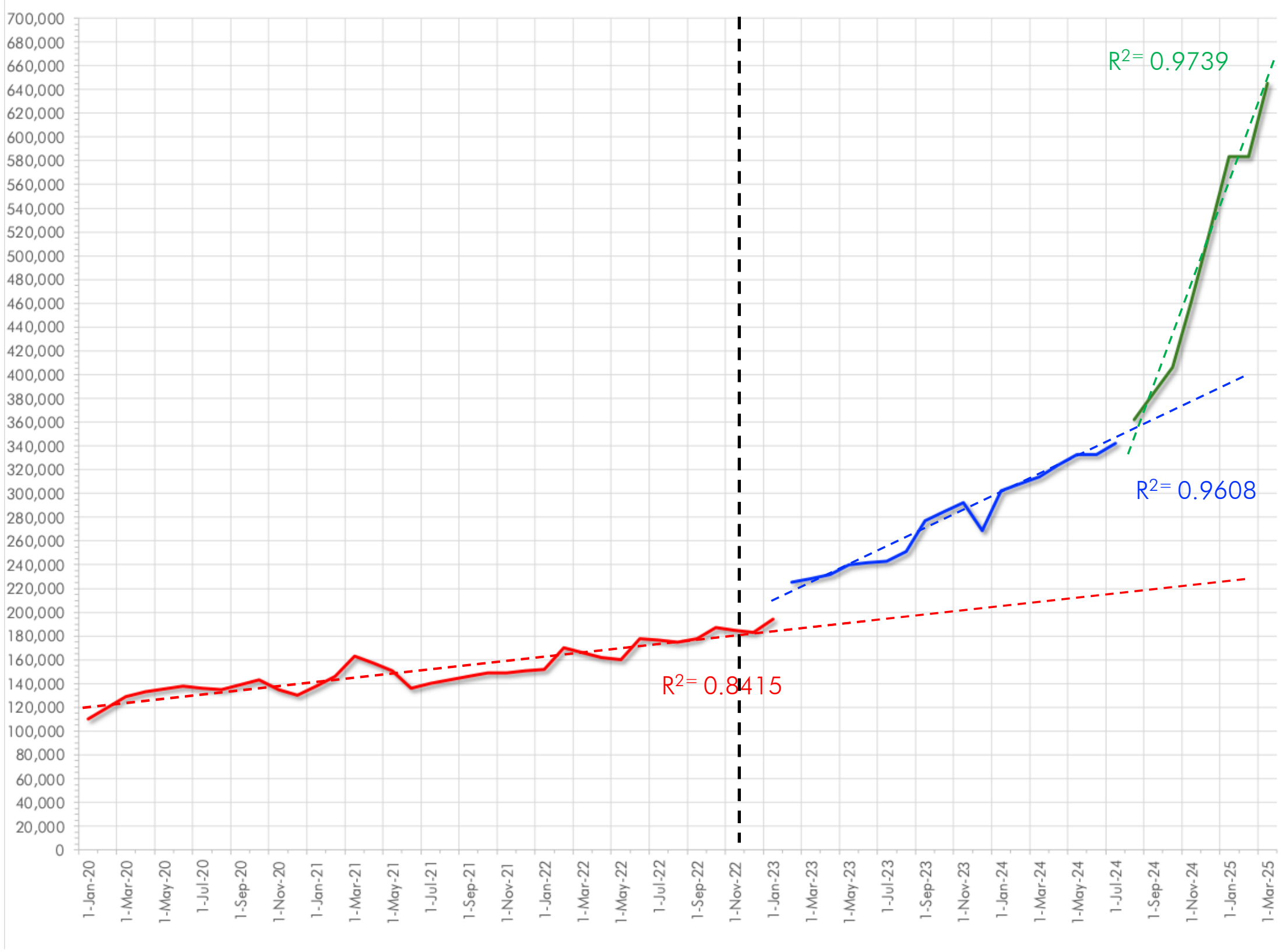

*Figure 7. Monthly frequency of webpages that contain the word 'explore' (in 1,000 for the period January 2020 to March 2025). The data have been smoothed by removing the troughs (see text). The dashed black line indicates the public release of ChatGPT in November 2022*

## References

[1]. Radford, A.; Narasimhan, K.; Salimans, T.; Sutskever, I. (2018). Improving Language Understanding by Generative Pre-Training. Available online: https://cdn.openai.com/research-covers/language-unsupervised/language_understanding_paper.pdf (accessed on

[2]. Brown, T.B.; Mann, B.; Ryder, N.; Subbiah, M.; Kaplan, J.; Dhariwal, P.; Neelakantan, A.; Shyam, P.; Sastry, G.; Askell, A., et al. (2020). Language Models are Few-Shot Learners [preprint]. 10.48550/arXiv.2005.14165. doi: 10.48550/arXiv.2005.14165.

[3]. OpenAI. (2023). ChatGPT 3.5 (August 3 version). Available online: https://chat.openai.com (accessed on Sep 11, 2023).

[4]. OpenAI. (2024). GPT-4 System Card. San Francisco CA: OpenAi.

[5]. OpenAI. (2024). GPT-4o System Card. San Francisco CA: OpenAi.

[6]. OpenAI. (2025). GPT-4.5 System Card. San Francisco CA: OpenAi.

[7]. Baack, S. (2024). A critical analysis of the largest source for generative ai training data: Common crawl. In Proceedings of Proceedings of the 2024 ACM Conference on Fairness, Accountability, and Transparency; pp. 2199-2208.

[8]. Common crawl. (2025). Common Crawl maintains a free, open repository of web crawl data that can be used by anyone. Beverly Hills, CA: The Common Crawl Foundation. Available online: https://commoncrawl.org (accessed on Mar 26, 2025).

[9]. Spennemann, D.H.R. (2025). The origins and veracity of references 'cited' by generative artificial intelligence applications. *Publications* 13 p. 12. doi: https://doi.org/10.3390/publications13010012.

[10]. Spennemann, D.H.R. (subm.). Pushing 150: How two generative Ai models understand time and envision the future. *Foresight* p. [under review]. doi.

[11]. Newsguard. (2025). Tracking AI-enabled Misinformation: 1,254 'Unreliable AI-Generated News' Websites (and Counting), Plus the Top False Narratives Generated by Artificial Intelligence Tools. Available online: https://www.newsguardtech.com/special-reports/ai-tracking-center/ (accessed on Mar 25, 2025).

[12]. Cyphert, A.B.; Blake, V.K. (2025). Code Blue: The Threat of Synthetic Data Use to Generative Medical AI. *Houston Journal of Health Law & Policy* 24 (1), pp. 167-190. doi.

[13]. Alemohammad, S.; Casco-Rodriguez, J.; Luzi, L.; Humayun, A.I.; Babaei, H.; LeJeune, D.; Siahkoohi, A.; Baraniuk, R.G. (2023). Self-consuming generative models go mad. *arXiv preprint arXiv:2307.01850* 4 p. 14. doi.

[14]. Yang, S.; Ali, M.A.; Yu, L.; Hu, L.; Wang, D. (2024). MONAL: Model autophagy analysis for modeling human-AI interactions. *arXiv preprint arXiv:2402.11271*. doi.

[15]. Alemohammad, S.; Humayun, A.I.; Agarwal, S.; Collomosse, J.; Baraniuk, R. (2024). Self-improving diffusion models with synthetic data. *arXiv preprint arXiv:2408.16333*. doi.

[16]. Gambetta, D.; Gezici, G.; Giannotti, F.; Pedreschi, D.; Knott, A.; Pappalardo, L. (2025). Characterizing Model Collapse in Large Language Models Using Semantic Networks and Next-Token Probability. 10.48550/arXiv.2410.12341. doi: 10.48550/arXiv.2410.12341.

[17]. Spennemann, D.H.R. (2023). Will the age of generative Artificial Intelligence become an age of public ignorance? *Preprint* 10.20944/preprints202309.1528.v1 pp. 1-12. doi: 10.20944/preprints202309.1528.v1

[18]. Cantor, M. (2023). Nearly 50 news websites are 'AI-generated', a study says. Would I be able to tell? *The Guardian*. Available online: https://www.theguardian.com/technology/2023/may/08/ai-generated-news-websites-study (accessed on Mar 26, 2025).

[19]. Mello, J.P. (2024). Copyleaks Study Finds Explosive Growth of AI Content on the Web. Available online: https://www.technewsworld.com/story/copyleaks-study-finds-explosive-growth-of-ai-content-on-the-web-179161.html (accessed on Mar 26, 2025).

[20]. Spennemann, D.H.R.; Biles, J.; Brown, L.; Ireland, M.F.; Longmore, L.; Singh, C.J.; Wallis, A.; Ward, C. (2024). ChatGPT giving advice on how to cheat in university assignments: How workable are its suggestions? *Interactive Technology and Smart Education* 21 (4), pp. 690-707. doi: 10.1108/ITSE-10-2023-0195.

[21]. Spennemann, D.H.R. (2023). ChatGPT and the generation of digitally born "knowledge": how does a generative AI language model interpret cultural heritage values? *Knowledge* 3 (3), pp. 480-512. doi: 10.3390/knowledge3030032

[22]. Google. (2025). GoogleTrends. "Delve into". Google Inc. Available online: https://trends.google.com/trends/explore?date=2022-01-01%202025-03-26&q=delve%20into&hl=en (accessed on 2025,

[23]. Google. (2025). GoogleTrends. ChatGPT. Google Inc. Available online: https://trends.google.com/trends/explore?date=2022-01-01%202025-03-26&q=chatGPT&hl=en (accessed on 2025,

[24]. Mortensen, O. (2024). How Many Users Does ChatGPT Have? Statistics & Facts (2025). Available online: https://seo.ai/blog/how-many-users-does-chatgpt-have (accessed on Mar 26, 2025).

[25]. Milmo, D. (2023). ChatGPT reaches 100 million users two months after launch. *The Guardian*. Available online: https://www.theguardian.com/technology/2023/feb/02/chatgpt-100-million-users-open-ai-fastest-growing-app (accessed on

[26]. Siteefy LLP. (2025). How Many Websites Are There in the World? London: Siteefy LLP. Available online: https://siteefy.com/how-many-websites-are-there/ (accessed on Mar 26, 2025).

[27]. de Kunder, M. (2025). The size of the World Wide Web (The Internet). Available online: https://www.worldwidewebsize.com (accessed on Mar 26, 2025).